\documentclass[11pt,a4paper,fleqn]{article}

\usepackage[tbtags]{amsmath}
\usepackage{amssymb}
\usepackage{mathrsfs}

\newcommand{\ud}{\,\mathrm{d}}

\newcommand{\SO}{\mathrm{SO}}

\newcommand{\lagdens}{\mathscr{L}}

\newcommand{\refcite}[1]{\cite{#1}}

\newcommand{\eq}[1]{Eq.~\!\eqref{#1}}

\title{\textbf{\Large Reply to ``Comment on {`Noncommutative gauge theories and Lorentz symmetry'}''}}

\author{
{\normalsize \textbf{Rabin Banerjee and Biswajit Chakraborty}}\\
{\small \textit{S.N.~Bose National Centre for Basic Sciences, JD Block, Sector 3, Salt Lake, Kolkata 700098, India}}\\
{\small \texttt{rabin@bose.res.in}, \texttt{biswajit@bose.res.in}}\\[1.0ex]
{\normalsize \textbf{Kuldeep Kumar}}\\
{\small \textit{Department of Physics, Panjab University, Chandigarh 160014, India}}\\
{\small \texttt{kuldeepk@pu.ac.in}}
}

\date{}

\begin{document}

\maketitle

\begin{abstract}
This is a reply to ``Comment on {`Noncommutative gauge theories and Lorentz symmetry,'}'' {\em Phys.~Rev.} D 77 (2008) 048701 by Alfredo Iorio.\par \bigskip \par
\centerline{\underline{Journal reference: \textit{Phys.~Rev.} D 77 (2008) 048702}}
\end{abstract}

\bigskip\par\bigskip


This is a reply to the Comment \cite{AI} on our paper. The author has criticised our earlier work without precisely pointing out the error. However we remain content with our analysis for the following reasons.

The main confusion of the author concerns a naive textbook application of Noether theorem to the problem at hand. There are two distinctions to be noted. First, such applications involve only scalar parameters rather than vector or tensor ones as is the case here. Second, and more importantly, textbook applications do not discuss actions that have  parameters that are not included in the configuration space. In that case the Noether procedure gets nontrivially modified as was shown in our paper \cite{Banerjee:2004ev}. We shall here reinforce this point by first discussing a simple example from particle mechanics and finally make the connection with the field theoretic models considered earlier \cite{Banerjee:2004ev}.

Consider a nonrelativistic particle of mass $m$ moving in three dimensions, subjected to a constant external force $\vec{F}$:
\begin{equation}\label{1}
m \ddot{x}_i = F_i.
\end{equation}
It follows immediately that the rate of change of its angular momentum $\vec{J} = \vec{x} \times \vec{p}$, with $\vec{p} = m \dot{\vec{x}}$ being the linear momentum, is precisely the applied torque $\vec{\tau} = \vec{x} \times \vec{F}$:
\begin{equation}\label{2}
\frac{\ud \vec{J}}{\ud t} = \vec{\tau}.
\end{equation} 
Now observe that \eq{1} follows from the following action:
\begin{equation}\label{3}
S = \int  L \ud t = \int \left( \tfrac{1}{2} m \dot{x}^2 + \vec{F} \cdot \vec{x} \right) \ud t.
\end{equation}
It is natural to expect that the force $F_i$ transforms covariantly under $\SO(3)$ rotation, so that \eq{1} has a covariant form. Correspondingly, the action \eqref{3} is \emph{invariant} under $\SO(3)$ rotation. But despite its rotational invariance, this action does not yield a conserved charge, which is the angular momentum $\vec{J}$ in this case; it rather satisfies \eq{2}, as mentioned above. Let us derive this from the action $S$ using Noether's approach and make some important and relevant observations on the way, by considering its response to an infinitesimal $\SO(3)$ rotation,
\begin{equation}\label{4}
x_i \rightarrow x_i' = x_i + \delta x_i,
\end{equation}   
where $\delta x_i = \omega_{ij} x_j$, $|\omega_{ij}| \ll 1$, $\omega_{ij}= - \omega_{ji}$. Under this transformation $F_i$ also undergoes the transformation
\begin{equation}\label{5}
F_i \rightarrow F_i' = F_i + \delta F_i, \qquad
\delta F_i = \omega_{ij} F_j.
\end{equation}
The invariance of the action $S$ under this transformation implies
\begin{equation}\label{6}
0 = \delta S = \int \ud t \left( m \dot x_i \delta \dot x_i + F_i \delta x_i +  x_i \delta F_i \right).
\end{equation} 
Note that it is important to consider $x_i \delta F_i$ term here. As we shall see later that this will play an important role.

Equation \eqref{6} can be rewritten, using the fact that $\delta \dot x_i = \frac{\ud}{\ud t} (\delta x_i)$, as
\begin{equation}\label{7}
0 =\delta S = \int \ud t \left[ \frac{\ud}{\ud t} (m \dot x_i \delta x_i) - (m \ddot{x}_i - F_i) \delta x_i + x_i \delta F_i \right].
\end{equation}
We can now get rid of the central term involving $\delta x_i$, using the equation of motion \eqref{1}, so that the on-shell version of \eq{7} becomes
\begin{equation}\label{8}
0 = \delta S = \int \ud t \left[ \frac{\ud}{\ud t} (m \dot{x}_i \delta x_i) + x_i \delta F_i \right].
\end{equation}
Before proceeding further, let us note at this stage that we cannot regard the force $F_i$ as an auxiliary variable belonging to the configuration space. Variables like $F_i$ should be interpreted as coordinates in an extended space
hidden behind the external dynamics. Correspondingly, any of the transformations in $F_i$, in particular the rotation, cannot be generated by a naive Posson bracketting with the angular momentum generator $J_i$:
\begin{equation}\label{9}
\{ F_i, J_j \} \ne \varepsilon_{ijk} F_k
\end{equation}
unlike
\begin{equation}\label{10}
\{ x_i, J_j \} = \{ x_i, \varepsilon_{jkm} x_k p_m \} = \varepsilon_{ijk} x_k.
\end{equation}
Poisson brackets involving variables like $F_i$  can be defined, but only in the extended space. Therefore, for a \emph{nondynamical} variable like $F_i$ forcing a requirement of `dynamical consistency' on it and thereby saying that $\vec{F}$ does not transform under rotation does not make sense. For other phase-space variables, of course, the requirement of dynamical consistency
\begin{equation}\label{11}
\delta \phi(x_i, p_i) = \{ \phi(x_i, p_i), G \}
\end{equation}
must be satisfied for any symmetry transformation generated by $G$, where the $\delta \phi$ on the left-hand side is the `algebraic transformation' defined in the Comment \cite{AI}.

Coming back to \eq{8} we can now introduce a vector $\vec{\omega} = \{\omega_k\}$, dual to the antisymmetric tensor $\omega_{ij}$, as
\begin{equation}\label{12}
\omega_k = \tfrac{1}{2} \varepsilon_{ijk} \omega_{ij}.
\end{equation}
This enables us to cast \eq{8} in the form
\begin{equation}\label{13}
0 = \int \ud t \omega_k \left[ -\frac{\ud}{\ud t}(\varepsilon_{ijk} x_i p_j) + \varepsilon_{kij} x_i F_j \right].
\end{equation}
Now the arbitrariness of $\omega_k$ readily yields
\begin{equation}\label{14}
\frac{\ud}{\ud t}(\varepsilon_{ijk} x_i p_j) = \varepsilon_{kij} x_i F_j
\end{equation}
which is nothing but \eq{2} in component form. We now make few comments.
\begin{enumerate}
\item
The \emph{covariant} forms of the Eqs.~\eqref{1} and \eqref{2} were ensured by the fact that we started with an \emph{invariant} action $S$ in \eq{3}.
\item
This exercise demonstrates that $\SO(3)$ invariance of $S$ does not yield a conserved angular momentum $\vec{J}$ anymore. This is clearly in contrast with the translational invariance of $S$, as the Lagrangian $L$ changes only by a total time-derivative:
\begin{gather}
\vec{x} \rightarrow \vec{x'} = \vec{x} + \vec{a}, \\
L \rightarrow L' = L + \vec{F} \cdot \vec{a} = L + \frac{\ud}{\ud t} \big[ (\vec{F}t) \cdot \vec{a} \big].
\end{gather}
(Note that the translational symmetry is not preserved in presence of a nonconstant force $\vec{F}$.) The corresponding conserved charge being ($m \dot{\vec{x}} - \vec{F} t$), as follows from \eq{1} by simple inspection.
\item
The nonconservation of angular momentum $\vec{J}$, despite having an $\SO(3)$ symmetry in the action \eqref{3}, is entirely due to the `transforming' $\vec{F}$ coming from $x_i \delta F_i$ term in \eq{6} which gives rise to the torque $\vec{\tau} = \vec{x} \times \vec{F}$. One can therefore identify \eq{2} or \eq{14} as the criterion for $\SO(3)$ symmetry. Further, $\vec{J}$ still generates rotation on all the phase-space variables ($\vec{x}, \vec{p}$) and functions thereof but \emph{not} on $F_i$, as mentioned earlier.
\item
If $F_i$ were not to transform, then clearly $S$ is no longer invariant. And even if we insist on $\delta S = 0$ under the transformation \eqref{4}, then clearly we shall meet with a contradiction, To see that, set $\delta F_i = 0$ in \eq{8}. Then in place of \eq{13} we have
\begin{equation}\label{15}
\int \ud t \omega_k \frac{\ud}{\ud t} (\varepsilon_{kij} x_i p_j) = 0
\end{equation}
yielding $\frac{\ud}{\ud t}(\vec{x} \times \vec{p}) = 0$ as the equation, which appears to be a modified criterion for $\SO(3)$ invariance of $S$ in presence of a \emph{nontransforming} $\vec{F}$. But this is clearly in conflict with \eq{2}. Besides, note that one cannot actually set $\delta S=0$ in the left hand side of \eq{8} to begin with as $S$ in \eq{3} is no longer invariant if $\delta F_i = 0$. This indicates that the system \eqref{3} respects $\SO(3)$ symmetry only in presence of a transforming $F_i$ as in \eq{5}.
\end{enumerate}

Having studied the symmetry aspects of this particle model, we now turn our attention to our field-theory model and try to reassess the comments made by Iorio in the Comment \cite{AI}.

First of all, the existing similarity between this simple model from particle mechanics, \eq{3}, with the toy model (Eq.~(23) in our paper \cite{Banerjee:2004ev})
\begin{equation}\label{16}
S = \int \ud^4 x \lagdens = - \int \ud^4 x \left( \tfrac{1}{4} F_{\mu\nu} F^{\mu\nu} + j^{\mu} A_{\mu} \right)
\end{equation}
or the first-order (in $\theta^{\alpha\beta}$) terminated effective commutative theory (Eq.~(76) in our paper \cite{Banerjee:2004ev})
\begin{equation}\label{17}
S = - \int \ud^4 x \left[ \tfrac{1}{4} F_{\mu\nu} F^{\mu\nu} + \theta^{\alpha\beta} \left( \tfrac{1}{2} F_{\mu\alpha} F_{\nu\beta} + \tfrac{1}{8} F_{\beta\alpha} F_{\mu\nu} \right) F^{\mu\nu} \right]
\end{equation}
should be obvious. The constant background fields like the vector field $j^{\mu}$ and the tensor field $\theta^{\mu\nu}$ transforming under (homogeneous) Lorentz transformations ensure Lorentz invariance of the actions. Both $j^{\mu}$ and the noncommutative parameter $\theta^{\mu\nu}$ here are the counterparts of the constant force $F_i$ introduced in \eq{3}. Sheer presence of these `transforming' external parameters gives rise to nonconserving angular momentum tensors in these two respective theories (Eqs.~(38) and (82) in our paper \cite{Banerjee:2004ev}):
\begin{gather}
\label{18}
\partial_{\mu} M^{\mu\lambda\rho} - A^{\lambda} j^{\rho} + A^{\rho} j^{\lambda} = 0, \\
\label{19}
\partial_{\mu} M^{\mu\lambda\rho} - {\theta^{\lambda}}_{\alpha} F_{\mu\nu} \left( F^{\mu\alpha} F^{\nu\rho} + \tfrac{1}{4} F^{\mu\nu} F^{\rho\alpha} \right) + {\theta^{\rho}}_{\alpha} F_{\mu\nu} \left( F^{\mu\alpha} F^{\nu\lambda} + \tfrac{1}{4} F^{\mu\nu} F^{\lambda\alpha} \right) = 0,
\end{gather}
as happens in the particle model, as we have seen already. These two results were obtained in an \textit{ab initio} computation carried out in \cite{Banerjee:2004ev}. The main point of derivation is that we had to consider the terms involving $\delta j^{\mu}$ and $\delta \theta^{\mu\nu}$ in these respective analyses \cite{Banerjee:2004ev} as we did earlier in \eq{6} for the particle model. In these kinds of situations the currents $M^{\mu\nu\lambda}$ will fail to satisfy the continuity equation in these field theoretical systems as certain components of $j^{\mu}$ or $\theta^{\mu\nu}$ can be readily related to identifiable appropriate external `torques' $\vec{\tau}$. For example, one can get the time-derivative of $J^i \equiv \frac{1}{2} \varepsilon^{ijk} \int \ud^3 x M^{0jk}$ from \eq{18} as,
\begin{equation}\label{20}
\vec{\tau} = \frac{\ud \vec{J}}{\ud t} = \vec{R} \times \vec{j},
\end{equation}
where $\vec{R} = \int \ud^3 x \vec{A}(x)$. This has indeed the same structural form as that of the particle model \eqref{2}, with $\vec{j}$ indeed playing the role of force $\vec{F}$. The same holds for the space components of the constant background field $P^\mu$ introduced in \eq{7} of the Comment \cite{AI}. Likewise one can also relate appropriate components of $\theta$ to $\vec{\tau}$ as well for the model \eqref{17}. We will not require the explicit form here.

All these analyses clearly show that angular momentum tensor will \emph{not} satisfy continuity equation in presence of such transforming additional constant parameters in the theory which act as constant background fields and give rise to torque on the system. In \refcite{Iorio:2001qy}, the variations of the parameters $j^{\mu}$ or $\theta^{\mu\nu}$ were not considered. Also, it was demanded that \cite{AI,Iorio:2001qy}
\begin{equation}\label{23}
\Delta_f j^{\mu} =  \{ j^\mu, Q \} = 0, \qquad
\Delta_f \theta^{\mu\nu} = \{ \theta^{\mu\nu}, Q \} = 0
\end{equation}
following from their requirement of dynamical consistency.

But as we have pointed out earlier the Poisson bracket of any \emph{nondynamical} variable like $F_i$, $j^{\mu}$ or $\theta^{\mu\nu}$ can possibly be defined only in an extended space in a more fundamental theory that goes beyond the usual phase space discussed here. For example, one can envisage a situation where the additional variables describing the extended space correspond to an appropriate external system which is `robust' enough, i.e. which is very weakly influenced by the dynamics of the original variables. One can then possibly construct the `total angular momentum,' taking into consideration the contribution of these
additional variables, which can generate transformations in $F_i$, $j^{\mu}$ or $\theta^{\mu\nu}$. In other words, the `algebraic' transformations of $j^\mu$ or $\theta^{\mu\nu}$ given by
\begin{equation}
\delta j^\mu = {\omega^{\mu}}{_\nu} j^\nu, \qquad
\delta \theta^{\mu\nu} = {\omega{^\mu}}{_\lambda} \theta^{\lambda\nu} - {\omega{^\nu}}{_\lambda} \theta^{\lambda\mu}
\end{equation}
cannot be generated by the angular momentum $M^{\mu\nu} \equiv \int \ud^3 x M^{0\mu\nu}$, obtained solely from the variables occuring within the theory, through a Poisson bracket like \eqref{23}, just as the transformation $\delta F_i$ \eqref{5} could not be generated by a naive Poisson bracket of $F_i$ with
the angular momentum operator.

The SUSY model considered in the Comment \cite{AI} cannot be compared to any of the above mentioned models, as the equation of motion for the $D$-field can be used to eliminate it from the Wess--Zumino model \eqref{11} in the Comment \cite{AI} to yield a \emph{meaningful} on-shell version \eqref{15} in \cite{AI}. In contrast, neither our $j^\mu$, nor author's $P^\mu$ nor $\theta^{\mu\nu}$ can be eliminated in this manner, as the entire theory collapses---as has been noted in the Comment \cite{AI} as well. Consequently, they cannot be regarded as variables in the configuration space and Euler--Lagrange equation of motion of these fields does not make any sense. On the contrary, $D$-field has some similarity with Lagrange multipliers which are counted in the configuration space variables and enforce meaningful constraints on the theory, like $A_0$ in the Maxwell theory. Thus we see that this procedure has to be implemented case by case and only in those models where it can be done consistently.

We conclude here by summarising that all we wanted to demonstrate in our paper \cite{Banerjee:2004ev} was that to preserve Lorentz invariance of these models, where $j^\mu$ or $\theta^{\mu\nu}$ can be regarded as constant background fields, it is necessary to consider a transforming $j^\mu$ or $\theta^{\mu\nu}$. This is manifest from the structure of the actions \eqref{16} and \eqref{17} themselves. We have also seen that despite the Lorentz invariance, the models do not admit angular momentum tensor $M^{\mu\nu\lambda}$ satisfying $\partial_\mu M^{\mu\nu\lambda} = 0$, as these background fields act like forces, thereby generating external torque on the system, as can be seen by considering spatial components $M^{0ij}$. Also one cannot impose the requirement of dynamical consistency on any background fields. 

Finally, we would like to mention that there are reasons from black hole physics to expect that the length scale determined by $|\theta^{\mu\nu}|$ has a lower bound \cite{Doplicher:1994tu}. It thus becomes necessary to demand $\theta^{\mu\nu}$ to be constant from other physical considerations. But as we have shown this cannot be reconciled with the usual Poincar\'e invariance. Nevertheless, it has been pointed out recently that a twisted Poincar\'e symmetry can be reconciled with constant $\theta^{\mu\nu}$ \cite{Chaichian:2004za}. However these issues were not discussed by us in \cite{Banerjee:2004ev}. 


{\raggedright

}


\end{document}